\documentclass[aps,prl,twocolumn,showpacs,superscriptaddress]{revtex4}
\usepackage{graphicx}
\newcommand{\ket}[1]{\mbox{$|#1\rangle$}}

\def\<{\begin{equation}}
\def\>{\end{equation}}

\begin{document}
\title{Photonic quantum transport in a nonlinear optical fiber}

\author{Mohammad Hafezi}  \affiliation{Physics Department,
    Harvard University, Cambridge, MA  02138}
\author{Darrick E. Chang}
\affiliation{Center for the Physics of Information and Institute
for Quantum Information, California Institute of Technology,
Pasadena, CA 91125}
\author{Vladimir Gritsev}
\affiliation{Physics Department,
    Harvard University, Cambridge, MA  02138}
    \affiliation
    {Physics Department, University of Fribourg, Chemin du Musee 3, 1700 Fribourg, Switzerland}
\author{Eugene Demler}
\affiliation{Physics Department,
    Harvard University, Cambridge, MA  02138}

\author{Mikhail Lukin}

\affiliation{Physics Department,
    Harvard University, Cambridge, MA  02138}

\begin{abstract}
We  theoretically  study the transmission of 
few-photon quantum fields through a strongly nonlinear optical medium. We develop a general approach to investigate non-equilibrium quantum transport of bosonic fields through a finite-size nonlinear medium and apply it to a recently demonstrated experimental system where cold atoms are loaded in a
hollow-core optical fiber. We show that when the interaction between photons is
effectively repulsive, the system acts as a
single-photon switch. In the case of attractive
interaction, the system can exhibit either anti-bunching or
bunching, associated with the resonant
excitation of bound states of photons by the input field. These effects can be observed by probing statistics of photons transmitted through the nonlinear fiber.
\end{abstract} 

\maketitle

%%% Introduction %%%%
Quantum dynamics of strongly correlated  systems far from equilibrium is a new frontier of many-body physics. Intriguing phenomena involving such dynamics have recently been observed in diverse physical systems ranging from ultra-cold atoms~\cite{Widera:2008,Roati:2008,Billy:2008,Zenesini:2009} to individual spins in semiconductors \cite{Cronenwett:1998,Zumbuhl:2004}. At the same time, recent experiments using ultra-cold atoms and optical photons~\cite{Bajcsy:2009,Schnorrberger:2009} have opened the door for studies of a novel form of quantum transport involving strongly correlated photons. It has been predicted that these systems can allow for surprising behavior, such as the dynamical creation of a ``crystal'' of photons~\cite{Chang:2008}. While many similarities exist between these photonic systems and condensed matter systems involving massive particles, the photonic systems also present a unique set of challenges. In particular, they do not thermalize, and are inherently open, driven systems, which highlights the need to develop novel techniques for analysis.

%%%%%%%%%%%%%%%%%%%%%%%%%%%%%%%%%%%%%
\begin{figure}[t]
\includegraphics[width=.45 \textwidth]{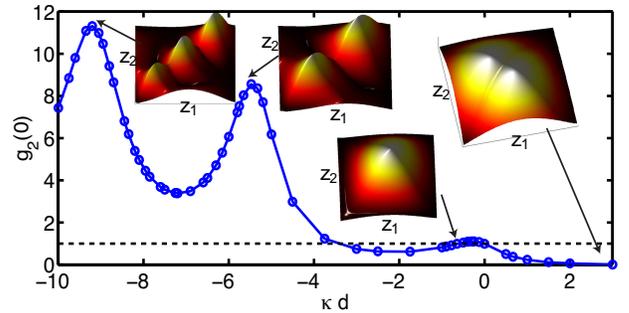}
\caption{Attractive ($\kappa<0$) and repulsive ($\kappa>0$) photons: Output
correlation function $g_2(\tau=0)$ as a function of nonlinearity.
The two-photon wavefunctions ($|\phi(z_1,z_2)|$) for four values of
the nonlinearity are shown. The system size is d=30.}
\label{fig:g2}
\end{figure}
%%%%%%%%%%%%%%%%%%%%%%%%%%%%%%%%%%%%%%

%%% Overview %%%

In this Letter, we describe a general technique to study the quantum transport of a few field quanta through a finite-length, strongly nonlinear one-dimensional waveguide. This technique allows us to determine the full spatial wavefunctions of the photons inside the waveguide as well as correlation functions of the outgoing reflected and transmitted light. We consider an optical waveguide in which the  tight confinement of photons near the diffraction limit \cite{ghosh:093902,kien:PRA2008,Bajcsy:2009} and the large number of atoms with which they interact should enable large optical nonlinearities at the single-photon level, which is necessary for applications like single-photon switching and photonic quantum gates~\cite{Kimble:2005, ShanhuiFan:PRL2007, Chang:2007}. Unlike nonlinear optical effects in cavity quantum electrodynamics~(QED)~\cite{Imamoglu:1997}, where only a single spatial cavity mode is involved, these waveguide systems are more difficult to treat in that they contain a large number of spatial degrees of freedom, much like low-dimensional, strongly interacting condensed matter systems~\cite{lieb-liniger,korepin:1993,Kinoshita:2004, Parades:2004,Caux:PRL2007} containing a few quanta. 
As a specific application, we use the calculated reflection
and transmission amplitudes to demonstrate how such a system can
be used to realize a single-photon switch. Our analysis reveals the tendency for photons to ``organize'' themselves due to an effective tunable  repulsive or attractive interaction [as shown in Fig.\ \ref{fig:g2}]. In the latter
case, we show that two-photon bound states can form inside the
waveguide, which can be observed using standard quantum optical
measurements of the output light.

%%%%%%%%%%%%%%%%%%%%%%%%%%%%%%%%%%%
\begin{figure}[b]
\includegraphics[width=.4 \textwidth]{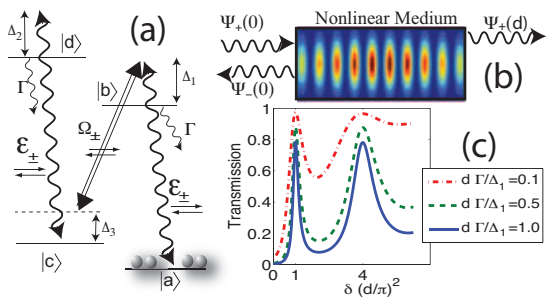}
\caption{(a) Four-level atomic
system yields a field evolution equation in the form of an NLSE.
(b) A schematic of quantum transport inside
a finite-length waveguide. An input field $\hat{\Psi}_{+}(0)$ is
injected into one end of the waveguide. Inside the medium, the
effective Bragg grating couples the forward- and
backward-propagating fields together and the evolution of the
coupled fields is described by the NLSE. A field is
transmitted~(reflected) at the end of the waveguide
$z=d$~($z=0$). (c) Linear transmission spectrum ($\kappa=0$) as a function of detuning, for a system of optical depth $\text{OD}{\equiv}n_0\text{L} \frac{\Gamma_{1D}}{\Gamma}=1000$. } \label{fig:four_level}
\end{figure}
%%%%%%%%%%%%%%%%%%%%%%%%%%%%%%%%%%%%%%

%%%%%% The Model %%%%%%%
Specifically, we are interested in the situation where the
dynamics of photons inside the waveguide is governed by the
quantum nonlinear Schr\"odinger equation~(NLSE),
\< i \frac{\partial {\hat{\Psi}}}{\partial
\tau}=-\frac{1}{2m}\frac{\partial^2 \hat{\Psi}}{\partial z^2}+2 \kappa \hat{\Psi}
^\dagger \hat{\Psi}^2.
\label{eq:NLSE}\>%
Here $\hat{\Psi}$ is a photonic field annihilation operator whose
specific form will be specified later. The presented framework can be used to treat the quantum transport of any NLSE system~(such as interacting atoms that are constrained to move along one dimension~\cite{Girardeau:1960,Kinoshita:2004,Parades:2004}), a problem that in general remains relatively unexplored. However, we focus on a photonic system implemented recently in Ref.\ \cite{Bajcsy:2009}, where an ensemble of atoms is loaded in a hollow-core fiber to mediate the interaction between photons. Moreover, we take advantage of quantum optical
techniques~\cite{Andre:bandgap,Bajcsy:2003} based on electromagnetically induced
transparency~(EIT) and photon trapping which allows us to dynamically control the effective mass
and interaction strength in Eq.~(\ref{eq:NLSE})~\cite{Chang:2008}. The light inside the fiber
interacts with an ensemble of atoms (with linear density $n_0$ and length $\text{L}$) whose level scheme is shown in
Fig.\ \ref{fig:four_level}. Following
Refs.~\cite{Bajcsy:2003,Chang:2008}, we use forward- and
backward-going dark-state polariton operators $\hat{\Psi}_{\pm}$,
which describe excitations of slowly-varying photonic fields
$\hat{\mathcal{E}}_{\pm}$ that are coupled to an atomic spin wave
excitation $\sigma_{ac}$ through a standing-wave control field
$\Omega_{\pm}(t)=\Omega$. The standing wave creates an effective Bragg grating that couples
$\hat{\Psi}_{\pm}$ together and traps them inside the
medium~\cite{Bajcsy:2003,andre:NLO}, much like an optical cavity.
In the limit of large optical depth ($\text{OD}=n_0\text{L} \frac{\Gamma_{1D}}{\Gamma}$), the symmetric combination
$\hat{\Psi}=(\frac{\hat{\Psi}_{+} + \hat{\Psi}_{-}}{\sqrt{2}})$
becomes the only independent quantity while the anti-symmetric
combination $\hat{A}=(\frac{\hat{\Psi}_{+} - \hat{\Psi}_{-}}{\sqrt{2}})$
adiabatically follows, $\hat{A}\simeq -\frac{i}{2m}
\partial_z \hat{\Psi}$. The detuning $\Delta_1$ leads
to a quadratic dispersion that can be interpreted as an effective
mass $m=\frac 1 2 (1+i\frac{\Gamma}{2|\Delta_1|})$. The presence
of the additional atomic state $\ket{d}$ leads to an optical
nonlinearity~\cite{Schmidt:1996} whose strength is given by
$\kappa=\frac{\Gamma_{1D}}{4(\Delta_2+i\Gamma/2)}$, where
$\Gamma_{1D}$ is the spontaneous emission rate into
the guided modes and $\Gamma$ is the total spontaneous emission
rate of both states $\ket{b}$ and
$\ket{d}$~($\Gamma_{1D}{\leq}\Gamma$). We are primarily interested
in the limit  $|\Delta_{1,2}|{\gg}\Gamma$ such that $m,\kappa$ are
mostly real and the evolution is dispersive.  Finally, we note that
Eq.~(\ref{eq:NLSE}) is written in dimensionless units; all lengths
and time are scaled by $\text{L}_{coh}=2 (\Delta_1^2+\Gamma^2/4)/\Gamma_{1D}n_0 |\Delta_1|$ and $\text{t}_{coh}=|\Delta_1|/2 \Omega^2$,
respectively.

%%%%% linear/nonlinear classical case %%%%%%
First, we investigative the linear transmission ($\kappa=0, \Psi=\langle\hat{\Psi}\rangle$)
spectrum of a finite system of dimensionless length $d=\frac{L}{L_{coh}}$, where an
input field $\Psi_{+}(0)$ is driving one end of the
system~[see Fig.~\ref{fig:four_level}(b)]. By solving a set of coupled mode equations for $\Psi_\pm$ \cite{andre:NLO}, one can show that like an optical cavity,
the coupling of $\Psi_{\pm}$ inside the waveguide creates a set of
transmission resonances in the output field $\Psi_{+}(d)$, which
due to the quadratic dispersion of the system occurs at
$\delta=(\frac{n \pi}{d})^2$ \cite{Hafezi:2009}. Here $\delta=\Delta_3 \text{t}_{coh}$ is the dimensionless
two-photon detuning and $n$ is a positive integer~[Fig.\
\ref{fig:four_level}(c)]. For a fixed resonant optical density
($\text{OD}\simeq d \frac{|\Delta_1|}{\Gamma} $), the system losses can
be tuned to yield either a sharp, low transmission peak~(large d)
or a high broad one~(small d) as shown in
Fig.\ref{fig:four_level}(c). The finesse of this effective cavity is
proportional to $\text{OD}$ \cite{Hafezi:2009}. For a classical field, addition of the
nonlinear term shifts the transmission peaks in frequency in an
intensity-dependent way to the left or right depending on the sign
of the nonlinearity coefficient. Transport in the classical NLSE system has already been extensively studied, where the nonlinearity is sufficiently weak that the operators in Eq.~\ref{eq:NLSE} can be replaced by complex numbers representing their mean-field values. In the context of atomic Bose-Einstein condensate transport~\cite{Jona:2003,Wimberger:2005,Korsch:2008}, this yields the well-known Gross-Pitaveskii equation, while in nonlinear optical fibers the classical NLSE explains, for instance, the formation of optical solitons~\cite{Agrawal:2007}. While the nonlinearity in our system can be made small enough to also observe such effects, here we are primarily interested in exploring the novel regime where the nonlinearity becomes extremely strong. In particular, when $\kappa \text{OD} \gg
1$, the frequency shift is predicted to be significant at
intensities corresponding to a single-photon level. In this
regime, the quantum transport of a few photons is expected to
behave fundamentally differently than predicted by classical
calculations, as we describe below.

%%%%% quantum formalism %%%%%%%%%%
In order to study quantum transport properties of the system, we
investigate the evolution of the wavefunction of the field in the
Schr\"odinger picture. We are interested in the situation where
the system contains no more than a few photons, which is
sufficient to describe the properties of a single-photon switch.
This allows us to truncate the Hilbert space so that only vacuum,
single-photon, and two-photon states are present~(the analysis can be
extended to truncating at
any small number of photons). We thus write the general state as:%
\begin{eqnarray}|\psi(t) \rangle&=& \int dz_1 dz_2 \phi(z_1,z_2,t)\hat{\Psi}^\dagger(z_1)
\hat{\Psi}^\dagger(z_2) |0\rangle \nonumber\\
&+&\int dz \theta(z,t) \hat{\Psi}^\dagger(z) |0\rangle +\epsilon |0\rangle.
\label{eq:state}\end{eqnarray}%
The first, second and third terms on the right correspond to the two-photon,
single-photon and vacuum components, respectively. Importantly,
directly solving for the two-photon wavefunction $\phi(z_1,z_2,t)$
allows us to characterize any non-trivial spatial order between
two photons in the system. The evolution of the two-photon wavefunction $\phi(z_1,z_2,t)$ under Eq.~(\ref{eq:NLSE}) is given by%
\begin{equation} i \partial_t\phi=-\frac{1}{2m}
\left( \partial^2_{z_1}+\partial^2_{z_2}\right)
\phi\nonumber +2 \kappa \phi\delta(z_1-z_2), \label{eq:evolution}
 \end{equation}
while the single-photon state evolves as a free, massive particle.
Generally, direct integration of such evolution equations should
be a viable approach for any transport problem involving only a
few photons~(or any other particles). A novel characteristic of photonic systems, however, is that light inherently can also enter and leave. Thus a remaining task is to formulate an ``input-output''
formalism that allows one to relate the wavefunctions inside the
waveguide~($0{\leq}z{\leq}d$) to the fields injected into and
propagating out from the system and any of their correlation
functions, which we describe below.

Since the NLSE Hamiltonian commutes with the field number operator
($\hat{\Psi}^\dagger \hat{\Psi}$),  manifolds with different field
quanta are decoupled from each other inside the medium. However,
the system is driven with an input field at $z=0$, and different
manifolds can be coupled at the boundaries. In particular, for a
classical input field, the boundary conditions correspond to that
of a coherent state, $ \hat{\Psi}_+(z=0) |\psi(t)\rangle  =
\alpha_{0}e^{-i\delta\tau} |\psi(t)\rangle$ where $\alpha_0$ is
the coherent state amplitude and $\delta$ the detuning. Since
$\hat{\Psi}_{+}$ annihilates a photon, the boundary condition relates
any manifold with $n$ photons to that with $n-1$. Using the fact
that the anti-symmetric combination $\hat{A}$ adiabatically follows
$\hat{\Psi}$, we can write
$\hat{\Psi}_\pm=\frac{1}{\sqrt{2}}(\hat{\Psi}\mp i\partial_z
\hat{\Psi})$~(here we consider the mass to be real). Therefore, using the multi-mode wavefunction of the system from Eq.\ (\ref{eq:state}),  the
boundary condition for $z=0$ reads $\phi(0,z_2,t)-i\partial_{z_1}
\phi(z_1,z_2,t)|_{z_1=0}=\frac{\alpha_0}{\sqrt{2}}e^{-i\delta\tau}\theta(z_2,t)$
and
$\theta(z=0,t)-i\partial_z\theta(z=0,t)=\sqrt{2}\alpha_{0}e^{-i\delta\tau}\epsilon$, and $\epsilon \simeq 1$. Similar boundary conditions apply at $z=d$
with an input $\alpha_0=0$. The intensity or any other field
correlation function can easily be obtained from the photon
wavefunctions. For example, the normalized second-order correlation function
of the output field $\hat{\Psi}_+(z=d)$ is given by $
g_2(0)=\frac{4\left|\phi(d,d)\right|^2}{ \left(
\left|\theta(d)\right|^2+4\int dz \left|\phi(z,d)\right|^2
\right)^2}\label{eq:g2_0}$.

%%% Two-photon spectrum %%%%%%
Before studying the dynamics, we analyze the
fundamental modes and energy spectrum supported by the system in
the absence of any input field ($\alpha_0=0$), which can be found
via the Bethe ansatz technique~\cite{lieb-liniger,Lai:1989b}. This ansatz
specifies that the eigenstates consist of a superposition of
states in which colliding particles exchange their wavenumbers
$k_i$. Unlike the typical formulation, the values of $k_{i}$ here
can be complex to reflect the \textit{open} nature of our boundary
conditions, which allow particles to freely enter or leave.
Imposing the boundary conditions, one finds that the fundamental
modes satisfy:
\begin{equation}
e^{2 i k_i d}=\frac{(k_i+1)^2}{(k_i-1)^2}\frac{(k_i-k_j+i\kappa)(k_i+k_j+i\kappa)}{(k_i-k_j-i\kappa)(k_i+k_j-i\kappa)}
\label{eq:momentum}
\end{equation}
where $(i,j)$ can be (1,2) and $i \neq j$ (for detailed derivation and generalization for many-body case see Ref.\ \cite{Hafezi:2009}). Correspondingly, the energy of each
state $E=k_1^2+k_2^2$ can acquire an imaginary part describing its
rate of exit. One class of allowed solutions corresponds to the
case where each photon is freely propagating. In the strong
interaction limit, these wavevectors exactly
match the allowed values for the non-interacting case, regardless
of whether the interaction is  attractive~($\kappa>0$) or
repulsive~($\kappa<0$). In the strongly repulsive case, this
phenomenon is well-known as the fermionization of hard-core
bosons~\cite{Girardeau:1960}. For a large system ($d\gg1$), the
non-interacting wavevectors and energy can be approximated by
$(\frac{m\pi}{d},\frac{n\pi}{d})$ and $\left(\frac{n \pi
}{d}\right)^2+ \left(\frac{m \pi }{d}\right)^2$, respectively. A
second class of solutions appears for negative $\kappa$ and
corresponds to bound states whose momenta for large system size
can be approximated by $(k_1,k_2) \simeq \frac{n\pi}{d}\pm i
\frac{\kappa}{2}$. In other words,  the interaction Hamiltonian
$V=-2|\kappa|\delta(z_1-z_2)$ admits a single bound state in the
relative coordinate of the two photons, while the allowed momenta
in the center-of-mass coordinates are quantized and determined by
the system size. This yields a discrete set of bound state
energies $E_n^b\simeq 2\left(\frac{n
\pi}{d}\right)^2-\frac{\kappa^2}{2}$.

%%%%%%%%%%%%%%%%%%%%%%%%%%%%%%%%%%%%%
\begin{figure}[b]
\includegraphics[width=.45 \textwidth]{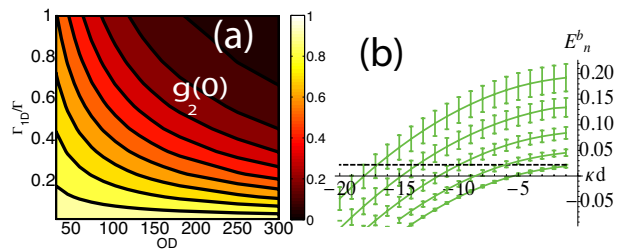}
\caption{(a) Repulsive photons: Correlation
function $g_2(\tau=0)$ of the transmitted light as a function of optical density ($\text{OD}=n_0\text{L} \frac{\Gamma_{1D}}{\Gamma}$) and single-atom cooperativity when the frequency is set to the single-photon transmission resonance with
$T\simeq 20\%$ and $\frac{\Delta_2}{\Gamma}=5$. (b) Attractive photons: Bound states energies (solid) become resonant with  energies of incoming photons (dashed) for specific values of $\kappa$. The error bars represent the decay rate of each bound state.}\label{fig:bound_g2_scaling}
\end{figure}
%%%%%%%%%%%%%%%%%%%%%%%%%%%%%%%%%%%%%

We now examine the properties of the light transmitted through the
waveguide, first considering the repulsive regime $\kappa>0$. We
fix the input field detuning to $\delta_{res}=(\pi/d)^2$, which
corresponds to the first transmission resonance in the linear
regime. Since we wish to study an open, driven system out of
equilibrium, conventional analytical methods such as Bethe ansatz
or quantum inverse scattering~\cite{korepin:1993} cannot be
applied here, whereas Eq.\ (\ref{eq:evolution}) with proper
boundary conditions can easily be numerically solved. Although all
numerical results correspond to a specific set of parameters
(system size, detuning, etc.), the conclusions are quite general.
In the presence of linear absorption, only a fraction of the
single-photon state is transmitted and the rest is dissipated
$(T<1)$; this results in the imperfect transmission shown in Fig.\ \ref{fig:four_level}(c),
since the single-photon state obeys linear dynamics. We note that since the system is weakly driven $(\alpha_0d)^2\ll1$, the noise effect due to absorption can be ignored \cite{Hafezi:2009}. On the other hand, the
two-photon state transmission is further suppressed due to an
extra nonlinear phase shift that brings this state out of
transmission resonance~(for a similar effect in a cavity, see
Ref.~\cite{Imamoglu:1997}), which
results in anti-bunching of the transmitted light~($g_{2}(0)<1$).
The anti-bunching implies that the transmitted field acquires non-classical character~\cite{Mandel:1986}, and the effect becomes more pronounced as $\text{OD}$ is increased,
which increases the effective system finesse [Fig.\
\ref{fig:bound_g2_scaling}(a)]. Physically, this behavior occurs because  transmission of the single-photon component of the field remains intact while two-photon and higher components become strongly reflected, and thus it can be said that this system operates as a single-photon switch. Moreover, we observe a significant
spatial deformation of the two-photon wave function inside the
waveguide as $\kappa$ is increased, as seen in Fig.\ \ref{fig:g2}.
In particular, a cusp develops along the diagonal $z_1=z_2$ and
the majority of the photon density lies in the off-diagonal
regions, indicating that the two photons repel each other. In the
limit of $\kappa{\rightarrow}{\infty}$, the diagonal elements must
be completely suppressed which leads to perfect anti-bunching. Similar behavior involving the ``self-organization'' of
photons in an NLSE system has been discussed in
Ref.~\cite{Chang:2008}.

%%%%  Attractive case  %%%%%%%
In the case of attractive interaction, the system exhibits a very
different behavior from the repulsive case.   In Fig.\
\ref{fig:g2}, we plot $g_{2}(0)$ for the transmitted field versus
${\kappa}d$. For small $|\kappa|d$, the nonlinear phase acquired by multi-photon components enhances~(suppresses) their reflection~(transmission), much like in the repulsive case. This is responsible for $g_{2}(0)$ dropping slightly below unity, despite the attractive interaction, and yields a weak single-photon switching effect. At larger values of $|\kappa|d$, oscillations develop in the correlation function, yielding
strong bunching behavior at particular values of ${\kappa}d$. A
closer analysis reveals that the locations of these resonances
correspond to resonant excitation of specific two-photon bound states,
$2\delta_{res}=E_{n}^{b}$. In particular, while the detuning $\delta_{res}$ of
the input field is fixed, the bound state energies $E_{n}^{b}$
vary quadratically with changing $\kappa$, causing the two to come
into resonance at particular values of $\kappa$. This is shown in
Fig.\ \ref{fig:bound_g2_scaling}(b), where the bound-state energies are
evaluated via Eq.~(\ref{eq:momentum}). This effect is further
confirmed by examining the two-photon wavefunction at each of
these oscillation peaks~[Fig.\ \ref{fig:g2}], where the photon
density becomes localized along the diagonal and indicates a bound
state in the relative coordinates. Simultaneously, an increasing
number of nodes and anti-nodes develop along the diagonal for
increasing $|\kappa|d$, which are associated with the higher
momenta of the center-of-mass motion. These resonances deviate
significantly from the predictions of the semiclassical picture
and also have no counterpart in cavity QED, due to the unique
spatial degrees of freedom available in our system. Importantly
for experiments, these bound states can be probed with classical
light, simply by examining higher-order correlation functions in
the output field. For example, the bunching peak associated with the second bound state at $|\kappa| d \simeq 5$, can be observed with parameter choices $T\simeq 1\%$
and $\frac{\Delta_2}{\Gamma}=-5$, using an optical density of $\text{OD}
\simeq 3500$ and coupling efficiency
$\frac{\Gamma_{1D}}{\Gamma}=0.2$.

In conclusion, we have presented a theoretical approach to investigate transport properties of a few field quanta inside a one-dimensional, finite-size nonlinear medium obeying the NLSE. When the system is driven out of equilibrium with a coherent field at the boundary, the strong interaction inside the medium is manifested in the correlation functions of the transmitted field. In particular, for a repulsive interaction, the transmitted field is anti-bunched, while for an effective attractive interaction, the transmitted field can be either bunched or anti-bunched. Moreover, in the attractive case, we show that the bunching is due to the resonant excitation of photonic bound states.  Recent progress in loading a large number of cold atoms into hollow-core optical fibers \cite{Bajcsy:2009} should allow one to probe these effects in ongoing experiments.

%%%  acknowledgements %%%%%%
We thank V. Gurarie, A. S\o rensen and S. Fan for useful discussions. This work was partially supported by  NSF, NSF DMR-0705472, Swiss NSF, CUA, DARPA, Packard Foundation and AFOSR-MURI. DEC acknowledges support from the Gordon and Betty Moore Foundation through Caltech's CPI, and NSF PHY-0803371.

\end{document}